\documentclass{sig-alternate}

\usepackage[utf8]{inputenc}
\usepackage[svgnames]{xcolor}
\usepackage{graphicx}
\usepackage{amsmath}
\usepackage{amssymb}
\usepackage{hyperref}
\usepackage{url}
\usepackage[algo2e,ruled,vlined]{algorithm2e}

\hypersetup{
    colorlinks, linkcolor={red!50!black},
    citecolor={blue}, urlcolor={red!50!black},
			pdfborder = {0 0 0}
}

\newcommand{\ShowTODO}[1]{{#1}}
\renewcommand{\ShowTODO}[1]{}

\newcommand{\TODO}[2]{\ShowTODO{\todo[inline, linecolor=#1, backgroundcolor=#1!60!white,bordercolor=#1]{#2}}}

\newcommand{\TODOYann}[1]{\TODO{gray}{{\bf TODO Yann :} #1}}

\newcommand{\Ab}{{\sf A}}
\newcommand{\Ub}{{\sf U}}
\newcommand{\Cb}{{\sf C}}
\newcommand{\Gb}{{\sf G}}

\newcommand{\AllBases}{{\cal B}}
\newcommand{\Trans}[2]{{\delta(#1,#2)}}

\newcommand{\ub}{\text{``{\tt .}''}}
\newcommand{\op}{\text{``{\tt (}''}}
\newcommand{\cp}{\text{``{\tt )}''}}

\newcommand{\Target}{S}
\newcommand{\Bases}[1]{\mathcal{C}_{#1}}
\newcommand{\Mand}{\mathcal{M}}
\newcommand{\Forb}{\mathcal{F}}
\newcommand{\Pairs}{\mathcal{P}}
\newcommand{\Rules}{\mathcal{R}}

\newcommand{\Gram}{\mathcal{G}}
\newcommand{\Aut}{\mathcal{A}}
\newcommand{\Lang}[1]{\mathcal{L}_{#1}}

\newcommand{\NTX}[3]{#2^{#1 \to #3}}
\let\chapter\section

\begin{document}
%
\conferenceinfo{BCB}{ '13, September 22 - 25, 2013, Washington, DC, USA}
\CopyrightYear{2013}
\crdata{978-1-4503-2434-2/13/09}


%
\title{Flexible RNA design under structure and sequence constraints using formal languages}

%
%
%
%
%

\numberofauthors{6} 
%
\author{
%
%
\alignauthor
Yu Zhou\titlenote{Both first authors equally contributed}\titlenote{Work was mainly performed while author was dually affiliated to: LRI, Univ. Paris-Sud, Orsay F-91405, France; and State Key Laboratory of Virology, Wuhan University, Wuhan, Hubei 430072, China. Current address: Department of Cellular and Molecular Medicine, Univ. of California, San Diego, La Jolla, CA 92093-0651, USA}\\
       \affaddr{LRI, Univ. Paris-Sud}\\
       \affaddr{Orsay F-91405, France.}\\
\alignauthor
Yann Ponty\raisebox{9pt}{$\ast$}\\
       \affaddr{LIX, Ecole Polytechnique}\\
       \affaddr{and AMIB, INRIA Saclay}\\
       \affaddr{Palaiseau F-91128,France}
\alignauthor St\'ephane Vialette\\
       \affaddr{LIGM, Univ. Paris-Est}\\
       \affaddr{Marne-la-Vall\'ee F-77454}\\
       \affaddr{France}
\and  
\alignauthor J\'er\^ome Waldispuhl\\
       \affaddr{McGill University}\\
       \affaddr{Montreal, Quebec H3A 2B2}
       \affaddr{Canada}
\alignauthor Yi Zhang\\
       \affaddr{State Key Laboratory of Virology,}
					\affaddr{Wuhan University,}\\
					\affaddr{and ABLife Inc,}\\
       \affaddr{Wuhan, Hubei 430072, China}
\alignauthor Alain Denise\\
       \affaddr{LRI and IGM, Univ. Paris-Sud}\\
       \affaddr{Orsay F-91405}\\
 					\affaddr{and AMIB, INRIA Saclay}\\
					\affaddr{France}
}

\ShowTODO{\setcounter{tocdepth}{1}
\listoftodos} 

\maketitle

\begin{abstract}
  The problem of RNA secondary structure design is the following: given a target secondary structure, one
  aims to create a sequence that folds into, or is compatible with, a given structure. 
  In several practical applications in biology, additional constraints must be
  taken into account, such as the presence/absence of regulatory motifs, either 
  at a specific location or anywhere in the sequence. 

  In this study, we investigate the design of RNA sequences from their targeted secondary structure, 
  given these additional sequence constraints. To this purpose, we develop a general framework based
  on concepts of language theory, namely context-free grammars and
  finite state automata. We efficiently combine a comprehensive set of constraints into a unifying context-free grammar of moderate size. From there, we use generic algorithms to perform 
  a (weighted) random generation, or an exhaustive enumeration, of candidate sequences. 
  
  The resulting method, whose complexity scales linearly with the length of the RNA, was implemented as a standalone program. The resulting software was embedded into a publicly available dedicated web server. The applicability of the method was demonstrated on a concrete case study dedicated to Exon Splicing Enhancers, in which our approach was successfully used in the design of \emph{in vitro} experiments.

\end{abstract}




\section{Introduction}

During the last years, the synthetic biology field and gene therapy techniques
have considerably  evolved \cite{Khalil:2010zr,Elowitz:2010ly}.
Ribonucleic acids (RNAs) emerged as versatile molecules capable to
serve as logic gates \cite{Win:2008ve} or to repress the replication of viruses
\cite{Ehsani:2010vn}. To perform their functions, these RNAs use specific structures
and nucleotide patterns enabling them to bind and interfere with other molecules.
Efficient methods for designing such molecules is thus key to reach the next milestones
in the emerging field of RNA nanotechnology~\cite{Guo:2010fk}.


In 1994, Hofacker and co-workers introduced the computational problem of RNA
secondary structure design (a.k.a. inverse folding)~\cite{HoFoSt1994} : Given a
secondary structure, design an RNA sequence (if any), that folds into the sought structure.
This definition reverses the classical RNA folding problem that consists in predicting
the secondary structure from sequence data. Unfortunately, while there is a polynomial
time and space solution for the folding problem without pseudo-knots~\cite{Zuker:1981fk},
the reciprocity remains questionable for the inverse folding problem~\cite{DBLP:conf/icml/Schnall-LevinCB08}.
Consequently, the Vienna group also proposed in their seminal 1994's paper a series
of creative heuristics for solving the RNA inverse folding problem, which have been implemented
in a program named \texttt{RNAinverse}~\cite{HoFoSt1994}.

In the late 2000s, the interest of the community for this problem grew with progresses and
potential applications in nano-biotechnologies. Thereby, Andronescu \textit{et al.} developed
\texttt{RNA-SSD}~\cite{AnFeHu2004},  Busch and Backofen \texttt{INFO-RNA}~\cite{BuBa2006,BuBa2007},
Zadeh \textit{et al.}~\texttt{NUPACK}~\cite{Zadeh:2011fk,Zadeh:2011uq}, Dai and Wiese
\texttt{rnaDesign}~\cite{DBLP:conf/cibcb/DaiTW09}, Levin \textit{et al.}
\texttt{RNA-ensign}~\cite{Levin:2012kx}, and Garcia-Martin \textit{et al.} \texttt{RNAiFold}~\cite{Garcia-Martin:2013zr}.
All forth mentioned programs improved over the original \texttt{RNAinverse}.

Soon though, a need to design RNA sequences with more complex structural properties as well as a necessity to
mimic properties of natural RNAs led to a second generation of software. First, the repertoire of designable shapes has
been extended with \texttt{Inv}~\cite{GaLiRe2010} and \texttt{fRNAkenstein}~\cite{Lyngso:2012vn} that can design 
sequences with (respectively) pseudo-knots or multiple alternate secondary structures. Next, methods have been proposed
to control generic sequence properties such as the genetic robustness with \texttt{RNAexinv}~\cite{Avihoo:2011ys}, or the
\texttt{GC}-content with \texttt{IncaRNAtion}~\cite{Reinharz2013}


Nevertheless, as mentioned earlier, the (secondary) structure alone is not sufficient
to fulfill all properties required to achieve most of RNA functions. In particular, our capacity to integrate multiple, precise and
highly specific sequence constraints is key to design functional RNAs. For instance, RNAs often require specific nucleotide
patterns to bind proteins (E.g. the hairpin of the TAR element).
Among existing methods, \texttt{INFO-RNA} is the first to enable its users to input sequence constraints, fixing a set of positions in the sequence where only some nucleotides are allowed. \texttt{NUPACK}~\cite{Zadeh:2011fk,Zadeh:2011uq} allows to prevent patterns in designed sequences, but cannot force the presence of a motif at an unspecific position.

Similar motif constraints (currently \Gb\Cb-content, positional constraints and a limit on runs of similar consecutive nucleotides) can also be defined with the recent \texttt{RNAiFold}~\cite{Garcia-Martin:2013zr}, arguably the closest competitor to our approach.
\TODOYann{Ajouter autres contraintes supportées par RNAiFold (nbres occ. de motifs), et citer Incarnation.}
However, the constraint programming approach used by the authors is not guaranteed to run in polynomial time and, in practice, can become prohibitively slow for longer sequences, or rich sets of constraints. 
Furthermore, the set of constraints supported by the software is rather restrictive, and a need for more general sequence constraints often arise in real applications. This is the case in~\cite{LiZhHu2010}, where RNA structures have been designed and tested {\it in vitro} in order to study the effects of RNA structures on the activity of exonic splicing enhancers (ESEs). In that work the sequential constraints were given by one mandatory motif (the ESE) at a fixed
position in the structure, and a (possibly large) set of forbidden motifs (other putative ESEs). Since such forbidden motifs could in
principle appear anywhere in the structure, they cannot be captured by positional constraints and may be hard to avoid for large sets of forbidden sequences.

\TODOYann{Reread once progress is made on the paper. JW: I emphasized the specificity of the approach. Please, check I'm correct.}
In this paper, we describe a general methodology for designing RNA
secondary structures subject to multiple positive and negative sequence constraints with or without location restrictions.
This methodology is based on concepts arising in formal language theory, namely context-free
grammars and finite automata, and on algorithms for the exhaustive or
random generation of words from a context-free language. In
Section~\ref{method} we describe our approach in its full generality. Then, in
Section~\ref{case}, we show how the method has been applied
for designing sequences that have been tested {\it in vitro} for a study on ESEs~\cite{LiZhHu2010}. 
Section~\ref{soft} describes our implementation, and a dedicated web server, offering a user-friendly environment for 
the generation and posterior analysis of sequences.
Finally we conclude in Section~\ref{conclusion} with some remarks and perspectives.

\section{Modeling constraints using language theoretical constructs}
\label{method}

It has been known for decades that concepts of language theory,
namely regular expressions and finite state automata (FSA) on one
hand, and context-free languages and grammars on the other hand, are
useful to the fields of genomics and bioinformatics.
Regular expressions and automata, and their probabilistic variants, as for example Markov models for
sequences, have been widely used for describing motifs in genomic
sequences.  Furthermore context-free grammars, again with probabilistic
variants like stochastic context-free grammars or covariance
models, have been used for structural bioinformatics, as a basis for
numerous works on prediction, comparison, and detection of RNA structures.

Here we combine these two formalisms (closely related
within the Chomsky/Schutzenberger hierarchy of formal languages) to address the problem of
designing RNA sequences that can fold to a given structure, while
respecting a set of constraints given by mandatory motifs (having fixed or variable position in the sequence)
and forbidden motifs.

\subsection{Method overview}

As stated above, our experimental applications require a rich combination of
constraints. Let $\AllBases := \{\Ab,\Cb,\Gb,\Ub\}$ be the set of nucleotides, i.e.
the vocabulary for our language constructs, these constraints can be broken down into the following four categories:
\begin{itemize}
\item {\bf Secondary structure constraint (CT):} Designed sequences are expected to
  be compatible with a target secondary structure $\Target$, i.e. base-pairing positions in the structure must correspond to canonical
base-pairs in the sequence (Watson-Crick \Gb--\Cb, \Ab--\Ub, or Wobble \Gb--\Ub).
\item {\bf Base positional constraint (CB):} Some positions within designed sequences must be chosen within position-specific restricted lists of bases.
In other words, each position $i$ is associated with a subset $\Bases{i} \subset \AllBases$. Note that such constraints can be adequately summarized by a sequence of IUPAC symbols, including the non-specific ${\sf N}\equiv \AllBases$.
\item {\bf Mandatory motifs constraint (CM):}  A predefined set $\Mand$ of sequences must be found {\bf at least once}, at an unspecified position, in any generated sequence.
\item {\bf Forbidden motifs constraint (CF):} Any designed sequence must avoid a predefined set $\Forb$ of forbidden sequences.
\end{itemize}
Notice that the length $n:=|S|$ of the designed sequence is implicitly set by the CB constraint.
Now, our problem can be formally stated as:
\begin{quote}\em Generate a set of $k$ sequences of length $n$, which are compatible with the target $\Target$ and $\Bases{}$, feature occurrences of each of the motifs in $\Mand$, and avoid the forbidden motifs in $\Forb$.
\end{quote}

\noindent Our method can be divided into four main steps:
\begin{enumerate}
\item Build a context-free grammar $\Gram$ that captures both the structure
  (CT) and base positional (CB) constraints, generating the language $\Lang{\Gram}$ of all sequences that are compatible with the base-pairing constraints induced by $\Target$, and the IUPAC sequence $\Bases{}$.
\item Build a deterministic finite-state automaton (DFA for short) $\Aut$
  which recognizes the language $\Lang{\Aut}$ of all sequences that
  feature a set $\Mand$ of motifs (CM), while avoiding a set $\Forb$ of motifs (CF).
\item Construct the intersection context-free grammar $\Gram_{\cap}$ that generates
  the language $\Lang{\Gram_{\cap}} := \Lang{\Aut} \cap \Lang{\Gram}$.
\item Use a (weighted) random generation algorithm to generate $k$ sequences from $\Gram_{\cap}$. Optionally, filter sequences satisfying additional properties, such as their affinity
towards the target structure, and other criteria, like the probability for certain regions to remain unpaired in the Boltzmann equilibrium.
\end{enumerate}
Note that $\Lang{\Gram_{\cap}}$ may be empty, as may result from a delicate interplay between the automaton and the grammar, if the constraints are too stringent.
This property, which may not be obvious from the grammar, can nevertheless be efficiently tested.




\subsection{Structural constraints and CFGs}

This first step consists in building a formal grammar, whose language
is exactly the set of sequences that are compatible with the input
structure and the set of positional constraints (CB).
Here, we assume that the secondary structure is free of pseudoknots, given as a dot-parenthesis notation and that corresponding parentheses are at least separated by one (unpaired) base ($\theta=1$, using standard nomenclature).

One builds a context-free grammar $\Gram=(\AllBases,{\cal V},\Rules,S)$, such that:
\begin{itemize}
\item $S:=V_1$ is the axiom.
\item ${\cal V} = \{V_i\}_{i\in \mathcal{X}}$, where $\mathcal{X}$ is the set of positions corresponding to an unpaired base, or to the 5' end of some base pair (denoted by an opening parenthesis in the dot-bracket notation).
\item The set of production rules is $\Rules=\bigcup_{i\in \mathcal{X}} \Rules_i$,
where $\Rules_i$ is a set of productions associated with each non-terminal $V_i$, and depends
on the status of the position $i$ in the target structure $\Target$.

For {\bf unpaired positions} ($\Target_i=\ub$), any nucleotide may be generated, provided that the positional constraint is satisfied. If there is no follow-up sequence at this level of nesting, then the generation stops here. Otherwise, the sequence must be extended, a task which is delegated to another non-terminal $V_{i+1}$.
\begin{align*}\Rules_{i} &:= \left\{\begin{array}{@{}c@{}l@{}}
\{V_i \rightarrow b\}_{b\in \Bases{i}} & \text{[If }\Target_{i+1}=\cp  \text{ or }i+1>n,\text{]}\\
\{V_i \rightarrow b\,V_{i+1}\}_{b\in \Bases{i}}& \text{[Otherwise.]} \end{array}\right.
\end{align*}

For {\bf paired positions} ($\Target_i=\op$, paired with $j$), the nucleotides chosen for the paired positions $(i,j)$ must belong to a list $\Pairs$ of valid base-pairs.
Again, there may or may not be a follow up sequence at the same level of nesting, leading to the following cases:
\begin{align*}\Rules_{i} &:= \left\{\begin{array}{cl}
\{V_i \rightarrow b\,V_{i+1}\,b'\;|\; (b,b')\in (\Bases{i}\times \Bases{j})\cap \Pairs\} \\
\text{[If }\Target_{j+1}=\cp \text{ or }j+1>n,\text{]}\\
\{V_i \rightarrow b\,V_{i+1}\,b'\,V_{j+1}\;|\;  (b,b')\in (\Bases{i}\times \Bases{j})\cap \Pairs\}\\ \text{[Otherwise.]}\\ \end{array}\right.
\end{align*}
\end{itemize}
In the above productions, we may typically allow only canonical base-pairs, thus:
$$\Pairs := \{(\Ab,\Ub),(\Cb,\Gb),(\Gb,\Cb),(\Gb,\Ub),(\Ub,\Ab),(\Ub,\Gb)\}.$$

\begin{figure}[t!]
  \includegraphics[width=\linewidth]{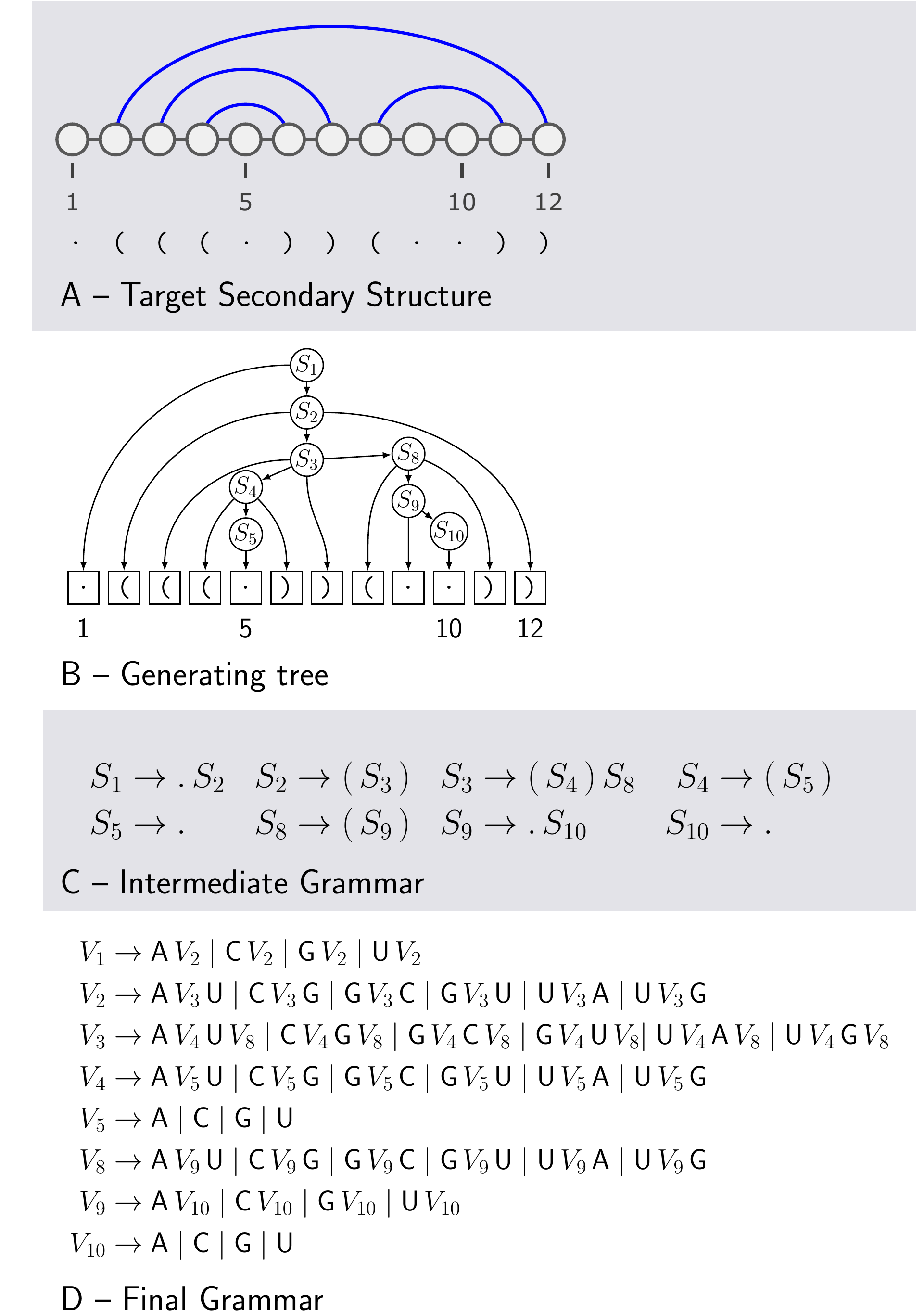}
\caption{Illustration of the grammar construct in the absence of positional constraints:
Any target secondary structure $\Target$ ({\sf A}) can be abstracted as a tree ({\sf B})
through a linear-time parsing, from which one derives an intermediate grammar ({\sf C})
whose sole production is $\Target$. The rules of this intermediate grammar can then
be duplicated to allow for alternative nucleotides, resulting into a final grammar
$\Gram$ ({\sf D}), which generates every RNA sequence compatible with $\Target$.
Additionally, the rules of $\Gram$ can be restricted to account for further positional constraints.\label{fig:exampleGrammar}}
\end{figure}
Note that these productions can be adapted in a straightforward manner from those of a
specialized grammar that only generates the targeted secondary structure.
The whole process of going from the secondary structure to the corresponding grammar
is illustrated by Figure~\ref{fig:exampleGrammar}.
The resulting grammar has $\Theta(n)$ rules, and can be generated in $\Theta(n)$ time and space
from the target secondary structure, given in dot-parenthesis notation, and a sequence of positional
constraints, given as a set of IUPAC symbols.

\subsection{Mandatory/forbidden motifs and automata}

We consider the language $\Lang{\Aut}$ of words that obey both the CM and CF constraints at the primary sequence level, and describe how to construct a deterministic finite-state automaton (DFA for short) $\cal A$ that recognizes $L_{\cal A}$. We assume some level of familiarity with the
basic notions and properties of finite-state automata, and direct the interested reader to Salomaa's classic reference~\cite{Salomaa1973}.

The existence of $\Lang{\Aut}$ follows from closure properties of regular languages through classical operations on sets (union, intersection, complement), and by concatenation. This means that the automaton that satisfies
our various constraints can be incrementally built from automata associated with each of the individual constraints.
However, the resulting automaton could become
too large thereby negatively impacting on the complexity of the whole method.
Therefore, we propose ad-hoc constructs based on the Aho-Corasick automaton, leading to much smaller automata.

\medskip
{\noindent\bf Formal language constructs.}
Let $\Lang{CM}$ and $\Lang{CF}$ be the languages of
words that respect the constraints $CM$ and $CF$, respectively.
Cleary, one has $\Lang{\Aut} = \Lang{CM} \cap \Lang{CF}$,
so it suffices to derive automata for $\Lang{CM}$ and $\Lang{CF}$.

%

Let us first consider the language $\Lang{CF}$ that contains all the words of $N^*$ which avoid all motifs in $\Forb=\{f_1,f_2,\ldots,f_{k'}\}$.
This language can be described as the complement in $N^*$ of $\Lang{\overline{CF}}$ the language that
asks for at least one occurrence of a motif in $\Forb$.
The complement language can be further described as the union $\Lang{\overline{CF}}= \bigcup_{i=1}^{k'}
\Lang{f_i}$, where $\Lang{f_i}$ is the set of words having at least one occurrence of $f_i\in\Forb$.
Clearly, $\Lang{f_i}$ is generated by the regular expression $N^*.f_i.N^*$, and hence
can be recognized by an automaton (having $|f_i|+1$ states).
Furthermore, regular languages are stable by union, so there exists an DFA that recognizes
$\Lang{\overline{CF}}$ and, taking the complement,
$\Lang{CF}$ is also recognizable by an automaton.

The general scheme for constructing a DFA that recognizes $\Lang{CM}$ is similar.
Let $\Mand=\{m_1, m_2, \ldots, m_{k''}\}$ be the set of mandatory motifs.
Clearly, one has $\Lang{CM} = \bigcap_{i=1}^{k''}
\Lang{m_i}$, where $\Lang{m_i}$ is the language of
all words containing at least one occurrence of $m_i$,
and can be constructed as described above.
Since regular languages are closed by intersection,
then $\Lang{CM}$ is also regular and can be recognized by a DFA.

\begin{figure}[t!]
  \includegraphics[width=\linewidth]{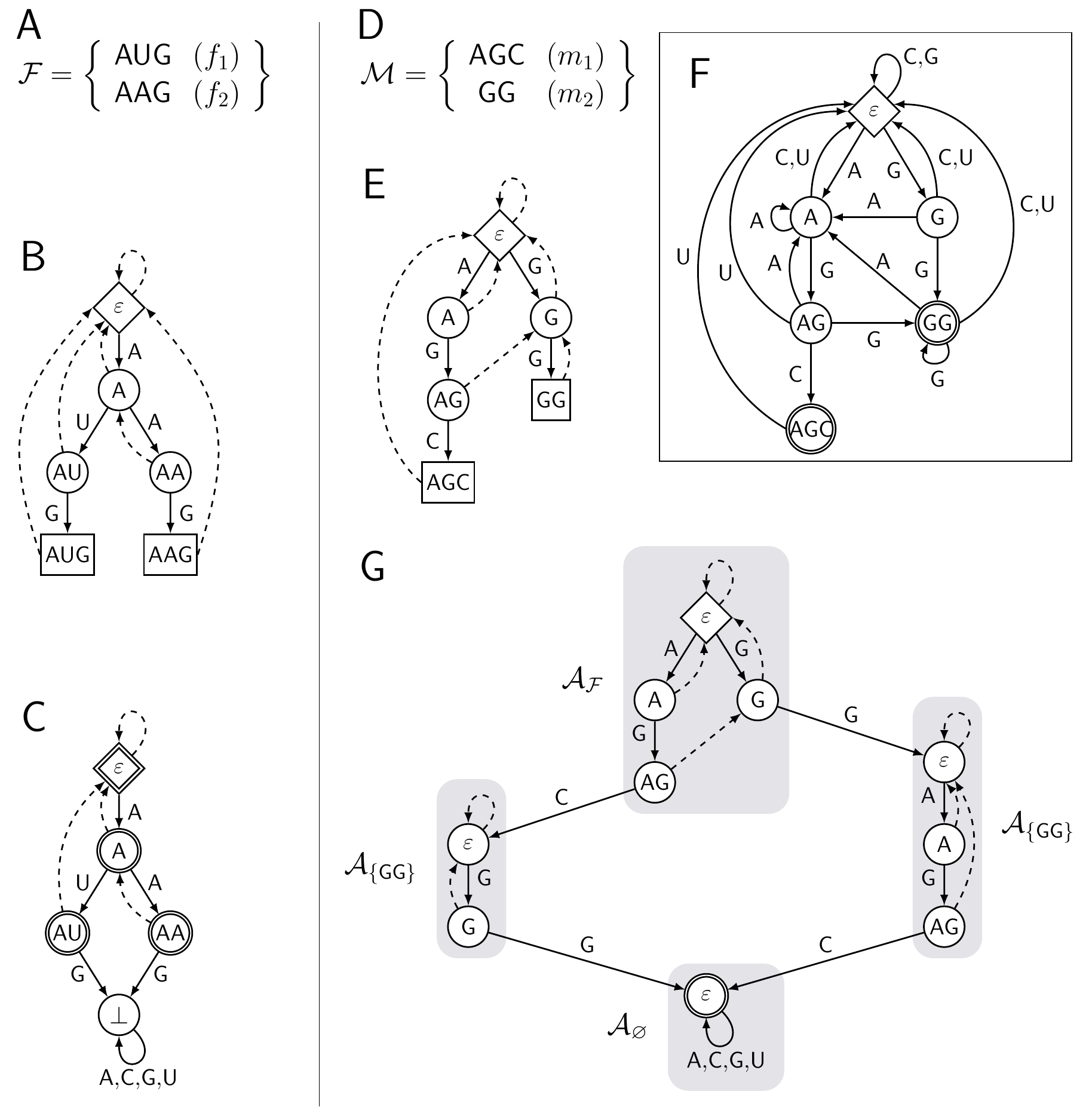}
\caption{Using the Aho-Corasick automaton to build compact automata for $\Lang{CF}$ and $\Lang{CM}$. From a set $\Forb$ ({\sf A}), an AC automaton can be built ({\sf B}) and transformed into an automaton for the language that excludes $\Forb$ ({\sf C}). For a set of mandatory motifs ({\sf D}), a similar initial construct can be used ({\sf E}), later to be duplicated to retain a memory of the motifs that are still to be generated ({\sf G}). The resulting automata can easily be  made deterministic and completed ({\sf F}).\label{fig:exampleACA}}
\end{figure}

\medskip
{\noindent\bf Aho-Corasick inspired automata.} The closure properties used by the above constructions
can be proven constructively, thus they can be used to build $\Aut$.
However, a naive implementation of these operations would yield a number of states in $\Theta((\sum_i |f_i|)\times \prod_{j} (1+|m_{j}|) )$. Such a potentially huge state space would lead to time-consuming computations while occupying a large amount of memory. However, it can be shown that a much smaller automaton exists for the same language.


Indeed, let us remind that automata for the language of words ending with some fixed motif can be adapted from the Aho-Corasick (AC) automaton~\cite{Aho1975}. As illustrated by Figure~\ref{fig:exampleACA}B and \ref{fig:exampleACA}E, an AC automaton is a simple prefix tree, complemented with failure transitions (dashed edges), which allows for an efficient, single-pass, string matching for a complete set of motifs ${\cal W}$. Namely, the AC automaton uses its states to represent the maximal prefix of a word in ${\cal W}$ which has been read/generated at any given moment.  Reading a new symbol either leads to a extended prefix, meaning a progression within the prefix tree, or a failure, resulting in a transition to a state associated with another smaller (possibly empty) prefix.
Note that, despite its classic representation as a non-deterministic automaton, the AC automaton can easily be made deterministic in linear time, as illustrated by Figure~\ref{fig:exampleACA}E and \ref{fig:exampleACA}F, increasing the number of edges (in $\Theta(\sum_i{|w_i|})$) at most by a constant factor $|\AllBases|$.

From such a construction, applied to the set $\Forb$ of forbidden motifs (cf Figure~\ref{fig:exampleACA}B), an automaton for $\Lang{CF}$ can be obtained by simply disallowing (rerouting to a non-accepting ground state $\bot$) each transition leading to a complete motif in $\Forb$. Any other state in the AC automaton is set as final/accepting, and the initial state is the one associated with the suffix $\varepsilon$, as illustrated by Figure~\ref{fig:exampleACA}C.

A similar approach can be used for $\Lang{CM}$. This time, one may simply create $2^{|\Mand|}$ versions of the AC automaton, each built for a subset of motifs $\Mand' \subseteq \Mand$ that remains to be generated. Reflecting these semantics, any transition leading to a complete motif $m_i$ in an automaton $\Mand'$ with be rerouted to a state, in the automaton $\Mand'':= \Mand'/\{m_i\}$, which corresponds to the longest suffix of $m_i$ which is a prefix of a word in $\Mand''$.
The initial state will be the one associated with the prefix $\varepsilon$ in the automaton $\Mand$, and any state in the automaton $\varnothing$ is accepting/final, as illustrated by Figure~\ref{fig:exampleACA}G.

These two constructs may even be merged into a single one (omitted from the example), that directly builds an automaton for $\Lang{CM} \cap \Lang{CF}$. One simply builds AC automata for each set of words $\Forb \cup \Mand'$, $\forall \Mand'\subseteq \Mand$. Any transition towards a state associated with a forbidden word $f\in\Forb$ is rerouted to the non-accepting ground state $\bot$ and any transition leading to a state $m\in\Mand$ in the automaton $\Mand'$ is rerouted to the automaton $\Mand'/\{m\}$ as described above. The initial state is then $(\varepsilon,\Mand)$, and any state $(-,\varnothing)$ is final. As summarized in Table~\ref{tab:states}, the resulting automaton has $\Theta(2^{|\Mand|}\cdot (\sum_i |f_i| + \sum_j |m_j|))$ states, and can be built in linear time. Note that the final automaton is not necessarily minimal, and one may obtain an even more compact automaton using any minimization algorithm.

\begin{table}[b]
\centering
  {\setlength{\tabcolsep}{3pt}
  \renewcommand{\arraystretch}{1.5}
  \begin{tabular}{| c | c | c |}
    \hline
    Lang. 
& \#States & Remark\\ \hline
    $\Lang{CF}$
    & $\le\displaystyle2+\sum_{i=1}^{|\Forb|} |f_i|$ & Aho-Corasick\\ \hline
    $\Lang{CM}$
    & $\displaystyle\le 2^{|\Mand|}\sum_{m_i \in \Mand} |m_i|$
    & {\begin{minipage}{9.5em}\small\centering Aho-Corasick \end{minipage}}\\ \hline
    $\Lang{\mathcal{A}}$
    & $\displaystyle\mathcal{O}\bigg(2^{|\Mand|}\sum_{i=1}^{|\Mand|} |m_i|\sum_{i=1}^{|\Forb|} |f_i|\bigg)$
    & Intersection\\ \hline
    $\Lang{\mathcal{A}}$
    & $\displaystyle\Theta\bigg(2^{|\Mand|}\cdot \big(\sum_i |f_i| + \sum_j |m_j|\big)\bigg)$
    & Aho-Corasick\\ \hline
  \end{tabular}}
  \caption{Number of states for the sequence constraint automaton.\label{tab:states}}
\end{table}




\subsection{Combining structural and sequential constraints}

It is well-known that context-free languages are not closed under intersection.
However, the intersection of a context-free language with a regular language is
always context-free.
The standard proof of the latter statement involves running a deterministic finite state
automaton in parallel with a pushdown automaton, and noting that the whole process can itself
be simulated by a pushdown automaton that accepts by final states.
Of particular interest in our context,
there exists an efficient algorithm that computes the context-free grammar
of the intersection of a context-free language with a regular language, as soon as the former
is given as a context-free grammar, and the latter is given as a
deterministic finite state automaton~\cite{Salomaa1973}
(the algorithm does not need to construct nor explicitly simulate a pushdown automaton).
In its original form, the algorithm needs the input grammar to be transformed into
an equivalent grammar in Chomsky Normal Form.
However, since our grammars have peculiar kinds of rules,
we can design a variant of the original algorithm that avoids this step.
Furthermore, this algorithm possibly generates useless rules and symbols, which do not generate any word due to conflicting constraints in both languages.
A {\em cleaning} step allows to remove these rules and symbols from the grammar.


The general principles of the algorithm are the following. We start
from a grammar $G=<N,V,P,S>$, where $N=\{\texttt{A}, \texttt{C}, \texttt{G}, \texttt{U}\}$ is the set of
terminal symbols, $V$ is the set of non-terminal symbols, $P$ is the
set of rules, and $S \in V$ is the axiom;
and from a DFA ${\cal A}=<N,Q,q_0,F,\delta>$,
where $Q$ is the set of states, $q_0 \in Q$
is the initial state, $F$ is the set of final states, and $\delta$ is
the transition function.
The output grammar $G'$ has its set of non-terminal symbols
$V' \subset Q \times V \times Q \cup S'$, where $S'$ is the axiom of $G'$.
Any symbol $(q,T,q') \in V'$ generates the words which are generated by $T$ in
the grammar $G$, and which correspond to a path from $q$ to $q'$ in
the automaton $\cal A$. Below are examples of this construction:
$$\begin{array}{@{}lcl@{}}
V_i \to b &\Rightarrow  &\{V_i^{q\to r}\to b\;|\; \Trans{q}{b}=r\}\\
V_i \to b\,V_{i+1} &\Rightarrow  &\{V_i^{q\to r}\to b\,V_{i+1}^{s\to r}\;|\; \Trans{q}{b}=s\}\\
V_i \rightarrow b\,V_{i+1}\,b' &\Rightarrow  &\{V_i^{q\to r}\to b\,V_{i+1}^{s\to t}\,b' \\
&&\quad\quad \;|\;\Trans{q}{b}=s\text{ and }\Trans{t}{b'}=r\}\\
V_i \rightarrow b\,V_{i+1}\,b'\,V_{j+1} &\Rightarrow  &\{V_i^{q\to r}\to b\,V_{i+1}^{s\to t}\,b'\,V_{j+1}^{u\to r} \\
&&\quad\quad \;|\;\Trans{q}{b}=s\text{ and }\Trans{t}{b'}=u\}\\
\end{array}$$

The algorithm begins by creating the axiom
$S'$ and one rule $S' \rightarrow [q_0,S,q]$ for every $q \in F$. A
stack allows to store every newly created non-terminal symbol. When a
symbol is popped out, all the rules starting with it are generated and
then pushed up.

Algorithm~\ref{algo_cfg_dfa} presents a basic implementation of these
principles (the easy cleaning step is omitted for the sake of simplification).
Additional technical improvements may be added (and are implemented in our
webserver) for the algorithm to become more space and time efficient,
but they are beyond the scope of this paper.

The algorithm produces a grammar having a set of rules in $\Theta(|\Rules|\cdot|Q|^3)$, each produced in constant average time. In the specific context of RNA rational design, let us remind that $|\Rules|\in\Theta(n)$ and $|Q|\in\mathcal{O}(2^{|\Mand|}\cdot (f+m))$ after minimization, where $f:=\sum_i |f_i|$ and $m:=\sum_j |m_j|$.
It follows that the overall time and space complexities of the complete algorithm grow like $\mathcal{O}(n\cdot 8^{|\Mand|}\cdot(f+m)^3)$, i.e. linearly on the sequence length.

\begin{algorithm2e}[t!]
  \label{algo_cfg_dfa}
  \SetKwData{base}{(\texttt{a}|\texttt{u}|\texttt{c}|\texttt{g})}
  \SetKwData{loops}{(aV|uV|..)}
  \SetKwData{stem}{(aVu|cVg|..)}
  \SetKwData{bifurcation}{(aVuW|cVgW|..)}
  \SetKwInOut{Input}{Input}
  \SetKwInOut{Output}{Output}
  \caption{Intersection of CFG with DFA}

  \Input{CFG $G=<N,V,P,S>$ and DFA ${\cal A}=<N,Q,q_0,F,\delta>$}
  \Output{$G'=<N,V',P',S'>$}
  \BlankLine

  \Begin{
    Create new axiom $S'$ \\
    $P' \leftarrow \varnothing$ \\
    $\text{NTqueue} \leftarrow \varnothing$ \\

    \ForEach{$q$ in $F$}{
      Add $\NTX{q_{0}}{S}{q}$ to \text{NTQueue} \\
      Add ($S' \rightarrow \NTX{q_{0}}{S}{q}$) to $P'$ \\
    }

    \While{$\text{NTqueue} \neq \varnothing$}{
      $\NTX{q}{V}{q'} \leftarrow \text{pop NTqueue}$ \\
      \For{$p \in P$ where $p.\text{leftHandSide} = V$}{
        \Switch{p}{
          \uCase(\tcp*[f]{Terminal \base}){$V \rightarrow u_{0}$}{
            $q' \leftarrow \delta\{q, u_{0}\}$ \\
            Add ($\NTX{q}{V}{q'} \rightarrow  u_{0}$) to $P'$ \\
          }
          \uCase(\tcp*[f]{\loops}){$V \rightarrow u_{0}V_{1}$}{
            $q_{1} \leftarrow \delta\{q, u_{0}\}$ \\
            Add ($\NTX{q}{V}{q'} \rightarrow  u_{0} \NTX{q_{1}}{V_{1}}{q'}$) to $P'$ \\
          }
          \uCase(\tcp*[f]{\stem}){$V \rightarrow u_{0}V_{1}u_{1}$}{
            $q_{1} \leftarrow \delta\{q, u_{0}\}$ \\
            \For{$q_{1}' \in Q$ where $\delta\{q_{1}', u_{1}\} = q'$}{
              Add ($\NTX{q}{V}{q'} \rightarrow  u_{0} \NTX{q_{1}}{V_{1}}{q_{1}'} u_{1}$) to $P'$ \\
            }
          }
          \uCase(\tcp*[f]{\bifurcation}){$V \rightarrow u_{0}V_{1}u_{1}V_{2}$}{
            $q_{1} \leftarrow \delta\{q, u_{0}\}$ \\
            \For{$q_{1}' \in Q$}{
              $q_{2} \leftarrow \delta\{q_{1}', u_{1}\}$ \\
              Add ($\NTX{q}{V}{q'} \rightarrow u_{0} \NTX{q_{1}}{V_{1}}{q_{1}'} u_{1} \NTX{q_{2}}{V_{2}}{q'}$) to $P'$ \\
            }
          }
          \uCase(\tcp*[f]{Only when V=S in current grammar}){$V \rightarrow V_{1}$}{
            \If{$q = q_{0}$ and $q' \in F$}{
              Add ($\NTX{q}{V}{q'} \rightarrow \NTX{q}{V_{1}}{q'}$) to $P'$ \\
            }
          }
          \Other(\tcp*[f]{No other cases in current grammar}){
            Exception \\
          }
        } 
      \lIf{NonTerminal in above productions is new}{
        Add it to NTqueue \\
      }
    } 
  } 
} 
\end{algorithm2e}


\subsection{Sequence generation}

From the final context-free grammar, a random generation of sequences can be done using
{\tt GenRGenS}~\cite{Ponty2006} or {\tt GrgFreqs} (a C++ re-implementation of {\tt GenRGenS}).
Both implementations feature  procedures for counting the total number of words compatible
with the constraints, their uniform random generation and their exhaustive enumerations.
These procedures rely on a recursive precomputation analogous to a
dynamic-programming scheme~\cite{Denise2010} for counting the number of words generated from
each non-terminal. From a weighted count, one can easily compute probabilities
associated with the choice of a given production rule among the ones accessible from a given non-terminal, in such
a way that applying the process recursively will result in a uniform random generation of
admissible sequences. A naive implementation of these procedures requires $\Theta(n^2.|\Rules|)$ arithmetic operations for general grammars. 

However, it can be remarked that the grammar $\Gram_{\cap}$ is rather peculiar, as each of its rules only produces
words of a single length. Consequently, the recursions used to compute the number of
generated sequences greatly simplify (as the convolution products responsible for the superlinear behavior are no longer justified).
They can consequently be computed in $\Theta(|\Rules|)$ time and $\Theta(|V|)$ space,
each generation requires $\Theta(|Q|\cdot n)$ arithmetic operations. {\tt GrgFreqs} has been extended to detect automatically such grammars, and adapt its algorithm accordingly.

\subsection{Filtering generated sequences}

Sequences generated from the grammar are
compatible with the target structure, and also satisfy all the
constraints on primary sequence. However, they may not necessarily fold into the MFE structure, i.e. admit the target secondary structure as their MFE.
To work around such a potential issue, we use the following approach:
\begin{itemize}
\item First, we use a weighting scheme to direct the random generation towards sequences of high affinity. Indeed, it was shown by some of the authors that using a weighted sampling 
strategy based on the free-energy greatly increases the probability of randomly 
generated sequences to fold into the target structure~\cite{Levin:2012kx,Reinharz2013}. Since {\tt GrgFreqs/GenRGenS} support such a non-uniform model~\cite{Ponty2006,Denise2010}, we included in our 
software an option to draw each sequence with respect to a pseudo Boltzmann-distribution. In this distribution, any admissible sequence is generated with probability 
proportional to $e^{-E/RT}$, where $R$ is the universal gas constant, $T$ is the temperature in Kelvin and $E$ is the free-energy of the target secondary-structure $\Target$. 
For the latter, we used a simplified version of the Turner 2004 energy model restricted to stacking base-pairs, requiring minor, yet technical, modifications of the grammar $\Gram_{\cap}$ 
(omitted in this presentation).
\item Then, a naive \emph{filtering} step selects sequences which
fold to the given structure and have high self-containment index~\cite{Lee2008}, probability to maintain the structure under its
sequence context. Other filters, like unpaired probability value
for a specific motif, can be applied to find better sequences
according to the requirement. 
\end{itemize}

For longer sequences, this strategy may still reveal insufficient to produce sequences that fold exactly into the target structure. 
In such situations, generated sequences may still be used as initial sequences (aka seeds) for traditional local search methods. 
Indeed, the combination of random generation (global sampling) and local search into a \emph{glocal} hybrid strategy was shown to outperform individual approaches, 
both in term of accuracy and diversity of the produced sequences~\cite{Reinharz2013}.



\section{Experiments}
\label{case}

We report here a case study that was done by some of the authors, using an early version of our software. 
This experimental study aimed to characterize how the structural context of an exon slicing enhancer (ESE) in a transcript affects its functionality~\cite{LiZhHu2010}. It is known that strong RNA secondary structures decrease the accessibility of the embedded exonic splicing enhancer (ESE) and thus hinder its recognition by single-stranded RNA binding proteins. This can result in enhanced or repressed exon splicing and different mRNA isoforms, which may have great impact on the functionalities of the proteins. In order to further characterize the effect of different RNA secondary structures on ESE functions, we needed to place the ESE motif in different structural context to test its functions experimentally. 
\begin{figure}[t!]
  \includegraphics[width=\linewidth]{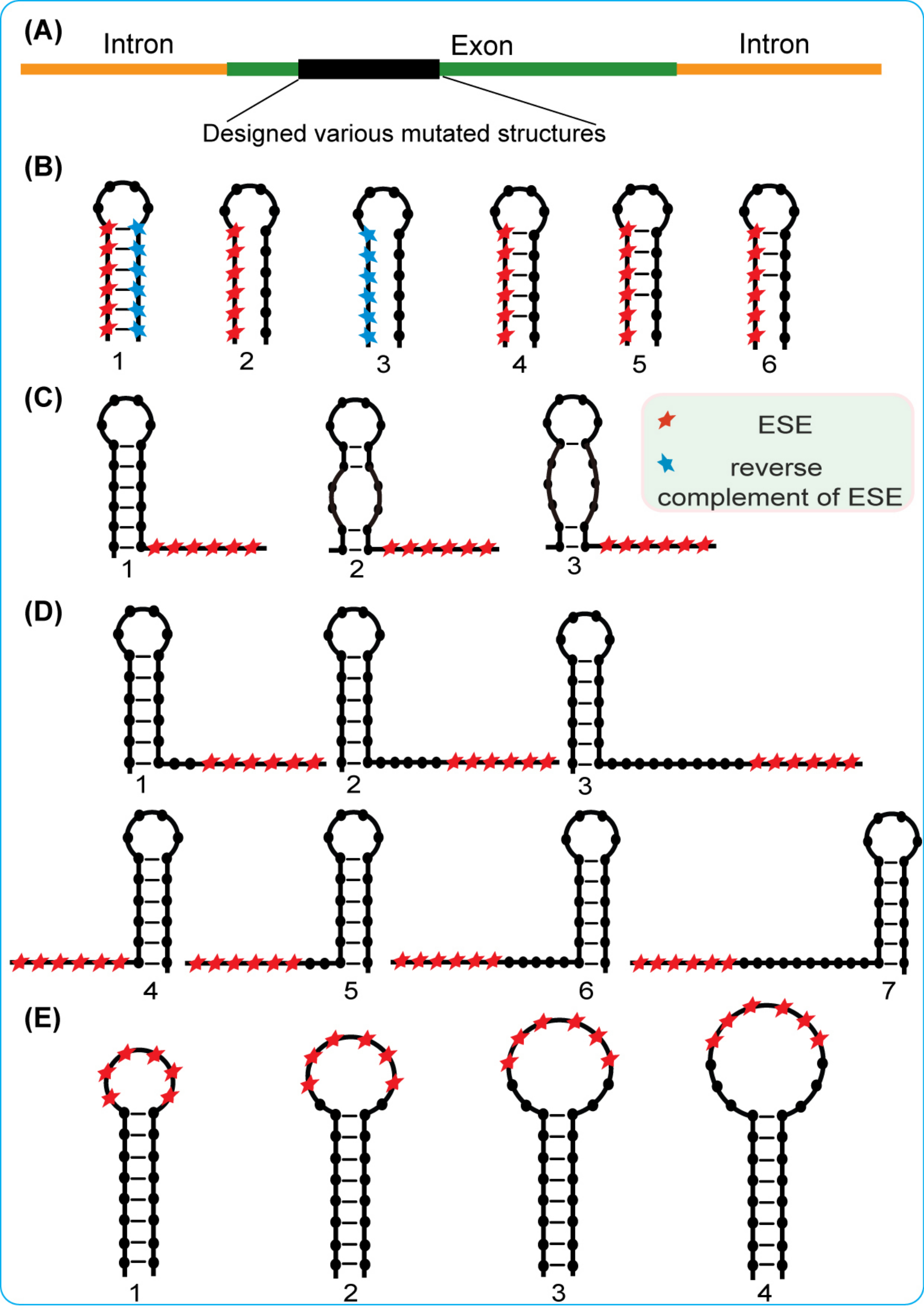}
\caption{Illustration of the experimental design (A) and various constructs containing the ESE motif in different structural context (B-D). ESE location are marked by the red star symbols. \label{fig:expdesign}}
\end{figure}

An {\it in vitro} splicing system was established by cloning the SMN1 splicing cassette into the GFP gene. The cassette sequence contained the SMN1 alternative exon 7 and its partial flanking introns. Various constructs (one ESE in different context) could be inserted into the exon (see Figure~\ref{fig:expdesign}A), and RT-PCRs were used to quantify the splicing outcomes reflecting the contextual effect on the ESE. 

Among the many known variants of the ESE motif~\cite{FaYeYe2004}, representative ESEs such as Tra2$\beta$-  and ASF-ESE) with high enhancer activities were selected to be tested. These variants were used as mandatory motif ({\bf CM}) in the constructs. We chose to consider a simple structure, a stem-loop, and to modify the structural context by putting it at different positions relatively to the ESE. More precisely, we designed four groups of constructs, including: 1) ESE in stem regions with varying stabilities by using different number of base-pairs (Figure~\ref{fig:expdesign}B); 2) ESE flanking by a varying stability stem (Figure~\ref{fig:expdesign}C); 3) ESE upstream or downstream a strong stem-loop with different lengths of spacers (Figure~\ref{fig:expdesign}D); 4) ESE in the varying-size loop of a strong stem-loop (Figure~\ref{fig:expdesign}E). The reverse complementary sequence of the ESE in Figure~\ref{fig:expdesign}B was used as an negative control ESE, leading to an experiment using another mandatory motif. 
Meanwhile, to avoid the confounding effects by other ESE motifs in the sequence, the constructs had to not contain any other potential ESE sequence ({\bf CF}), nor even any ESS sequence (Exonic Splicing Silencer). This set of forbidden motifs could be up to 1475 out of 4096 hexamers, if all the computationally identified motifs were used~\cite{FaYeYe2004}.

Starting from these constraints and using our methodology, we generated a set of candidates, further restricted by our experimental collaborators, later to be constructed and experimentally tested. The results  further supported that stem structure could block embedded ESE function, and showed that ESEs following a stable hairpin structure could be highly active. It was also found that different ESEs in the loop may function differently because of potential non-canonical base-pairings. This suggested that the  modulation on ESE functions by RNA structure could be prevalent and in different modes of actions. More details regarding the experimental aspects of the study can be found in Li~\emph{et al}~\cite{LiZhHu2010}.

\begin{table*}
{\scriptsize 
\begin{tabular}{|c|c|c|c|c|c|c|c|c|c|c|c|c|c|}
\hline
Test & Sequence/Structure & M            & \#F & \#state & \#trans & GI          & GA        & GB      & \#seqAG \\
\hline
1    & 
\begin{tabular}{c}
{\tt CUCGAACGCAANNNNNNNNNNAAUUC}\\
{\tt .....((((((....)))))).....}\\
\end{tabular}
                          & {\tt ACGCAA} & 1   & 902     & 3062    & 14142,17522 & 2010,6405 & 630,803 & 346 \\
\hline
2    & 
\begin{tabular}{c}
{\tt CUCGAACGCAANNNNNNNNNNAAUUC}\\
{\tt ..........................}\\
\end{tabular}
                          & {\tt ACGCAA} & 1   & 902     & 3062    & 906, 3064 & 903, 3064 & 903, 3064 & 457647 \\
\hline
3    & 
\begin{tabular}{c}
{\tt CUCGAACGCAANNNNNNNNNNAAUUC}\\
{\tt ......(((((....)))))......}\\
\end{tabular}
                          & {\tt ACGCAA} & 1   & 902     & 3062    & 24220, 29483 & 3123, 10438 & 636, 820& 1038 \\
\hline
4    & 
\begin{tabular}{c}
{\tt CUCGAACGCAANNNNNNNNNNAAUUC}\\
{\tt .......((((....)))).......}\\
\end{tabular}
                          & {\tt ACGCAA} & 1   & 902     & 3062    & 38507, 48783 & 5434, 18910 & 645, 856 & 4844 \\
\hline
5    & 
\begin{tabular}{c}
{\tt CUCGAACGCAANNNNNNNNNNAAUUC}\\
{\tt ........(((....)))........}\\
\end{tabular}
                          & {\tt ACGCAA} & 1   & 902     & 3062    & 48087, 56849 & 5826, 20167 & 663, 914 & 7958 \\
\hline
6    & 
\begin{tabular}{c}
{\tt CUCGANNNNNNUACAGANNNNNNAAUUC}\\
{\tt .....((((((......)))))).....}\\
\end{tabular}
                          & {\tt UACAGA} & 1   & 222     & 712    & 16643, 47520 & 14946,45655 & 1998, 5005 & 13948 \\
\hline
7    & 
\begin{tabular}{c}
{\tt CUCGANNNNNNNNNNNNNNNNUACAGAAAUUC}\\
{\tt .....((((((....))))))...........}\\
\end{tabular}
                          & {\tt UACAGA} & 0   & 0     & 0    & - & - & - & 0 \\
\hline
8    & 
\begin{tabular}{c}
{\tt NNNNNNNNNNUCGUCG}\\
{\tt (((....)))......}\\
\end{tabular}
                          & {\tt UCGUCG} & 1   & 1117     & 3889    
 & \begin{tabular}{c} 716553,\\1710596\\ \end{tabular}
 & \begin{tabular}{c} 219601,\\885384\\ \end{tabular}
 & 18223,34324
 & 42234 \\
\hline
9    & 
\begin{tabular}{c}
{\tt UCGUCGNNNNNNNNNN}\\
{\tt ......(((....)))}\\
\end{tabular}
                          & {\tt UCGUCG} & 1   & 939     & 3209    
 & \begin{tabular}{c} 115152,\\252205\\ \end{tabular}
 & \begin{tabular}{c} 32620,\\122590\\ \end{tabular}
 & 11469,20532
 & 35209 \\
\hline

\end{tabular}
}
\caption{Statistics of the RNA design on test data.  All the 9 tests use the same set of forbidden motifs which contains 238 putative hexamer ESEs. Sequence: expected sequence, {\sf N} means [\Ab\Ub\Cb\Gb]; Structure: bracket notation, `()' means pairing; M: Mandatory motif, present only once. Forbidden: forbidden motif set, in which the motifs cannot be present; \#F, \#state and \#trans means the number of final states, total states, and transitions, independently; GI, GA, GB represent the (numbers of non-terminal states, number of productions) pair statistics for the grammar after Initial intersection, after removing productions containing non-accessible non-terminals, and after clean the productions, independently; \#seqAG: number of sequences under both constraints of motifs and structure. \label{fig:experiments}}
\end{table*}


%

\section{Software and web server}
\label{soft}

The package named CFGRNAD, was mainly written in Python, and 
integrates the {\tt OCaML} {\tt FSA} program 
(\url{http://www.linguistics.ucla.edu/people/grads/jheinz/software/}), and
{\tt GrqFreqs}~\cite{Ponty2006}.  The package including the programs can be downloaded from \url{https://code.google.com/p/cfgrnad/}.

\subsection{Web server}

{\tt CFGRNAD} was embedded  in a web server available at:

{\centering\url{http://www.lix.polytechnique.fr/RNADesignStudio/}\\}

\noindent The server enables a user to submit a design task, specifying various constraints and options, such as:
\begin{itemize}
  \item Targeted secondary structure in dot-bracket notation;
  \item Position-specific constraints denoted by a sequence of IUPAC codes;
  \item Lists of forbidden and mandatory motifs, either manually specified or uploaded as a file;
  \item Number of generated sequences;
  \item Random generation model, chosen between a uniform model and a Boltzmann/weighted model.
  The latter weighs each eligible sequence with a probability $p \propto e^{-E/RT}$, where $E$
  is the sequence free-energy assuming the targeted structure, $R$ is the universal gas constant and $T$
  is the temperature in Kelvin. A simplified version of the 2004 Turner model for the free-energy is used, assigning
  individual contributions to each stacking pairs and disregarding other terms.
\end{itemize}

After computation, the resulting set of sequences is displayed and can be further analyzed within a dedicated web page, accessible at a later
time using a unique accession id. In this web space, the structure/sequence can be visualized using the VARNA software~\cite{VARNA} and each candidate
sequence can be evaluated using various statistics, including: 
\begin{itemize}
\item \Gb+\Cb-content; 
\item Free-energy of sequence upon forming the targeted structure, evaluated within the Turner model; 
\item Boltzmann probability of the targeted structure; 
\item Difference between the minimal free-energy structure and its second best suboptimal; 
\item Whether or not the targeted structure is the MFE for the candidate sequence. 
\end{itemize}
These statistics are computed on-demand from the result page using software
from the Vienna RNA software package~\cite{HoFoSt1994} (with the exception of the straightforward \Gb\Cb-content).

\begin{figure}[h]
\includegraphics[width=\linewidth]{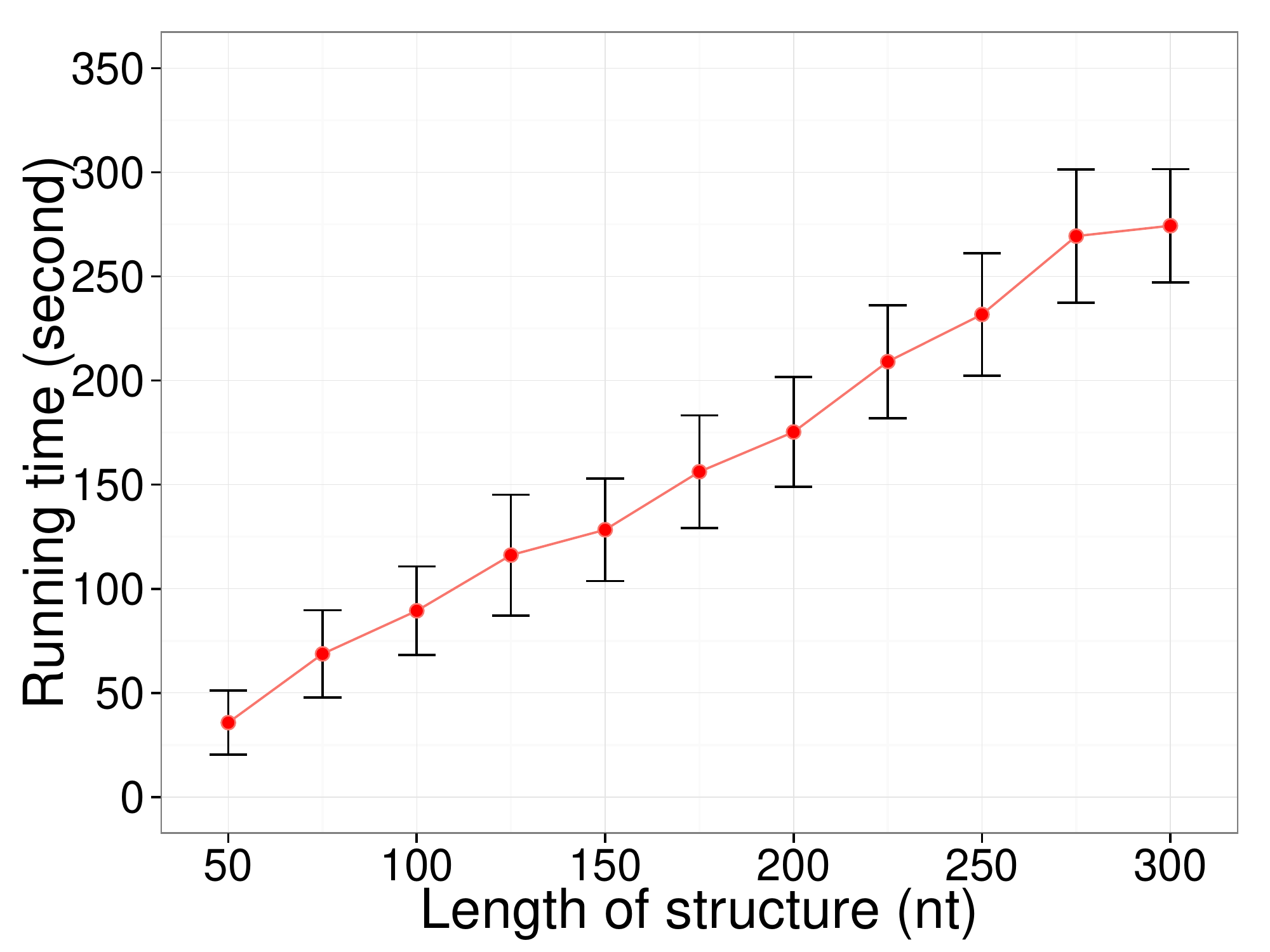}
\caption{Typical runtime on random realistic secondary structures and sets of motif $\Mand=\{\Cb\Gb\Ub\}$ and forbidden motifs $\Forb=\{\Ab\Ab\Ub,\Cb\Gb\Cb,\Ub\Gb\Cb\}$.\label{fig:runtime}}
\end{figure}

\subsection{Runtime and robustness testing}

To confirm its linear complexity, we measured the runtime of our random generation software. To that purpose, we used a realistic random model used by Levin~\emph{et al}~\cite{Levin:2012kx}, which performs a weighted random generation to draw structures which, on the average, have similar structural properties (\#base-pairs/length, \#helices/length\ldots) as experimentally-determined ribosomal RNAs.
For length varying from 50 to 300nts by steps of 25nts, we generated 20 random secondary structures. For each of the generated structure, we designed 100 candidates, using a single mandatory motif $\Mand=\{\Cb\Gb\Ub\}$ and three arbitrary forbidden motifs $\Forb=\{\Ab\Ab\Ub,\Cb\Gb\Cb,\Ub\Gb\Cb\}$. The results, summarized in Figure~\ref{fig:runtime} by averaging the runtime of sequences of equal length, confirm the linear complexity of the software, while exhibiting a fairly low variance.

We also performed more involved tests of our implementation on experimentally-determined structures, leading to statistics are summarized in Table~\ref{fig:experiments}. For all these tests, we used a much larger set of forbidden motifs proposed by  Fairbrother \emph{et al}~\cite{FaYeYe2004}, consisting of 238 putative ESEs. We noticed that the size of FSA became much larger than that on a small dataset, which contained only 4 or 5 2-mers, and the time required to build the FSA also increased. The cost of the intersection between the grammar and the FSA increases with the size of the automaton. 
Test 8 has the highest run time for intersection, the reason for which is probably that both side of the stem do not have motif base constraints. It is interesting that the polarity of the structure influences the time of intersection, as can be noticed by comparing test 8 to test 9. One of the reasons may be that the stem on the 5' end in test 8 generates too much unused non-terminals, or productions, than the stem at the 3' end in test 9. Finally, test 7 reveals that the list of putative ESEs conflicts with the mandatory constraints, revealing an unfeasible design objective. Being able to identify such situation is one of the strength of our method, especially in comparison with local search alternatives.

\subsection{Potential for inverse folding}
As emphasized in our introduction, our approach focuses on positive design principles (target + constraints compatibility), but does not explicitly capture negative design goals (specificity, avoidance of more favorable folds). However, its low complexity allows for a generation of a large number of independent candidates, which can be refolded and tested. A first benchmark was proposed in Reinharz \emph{et al}~\cite{Reinharz2013}, with a strong emphasis on the {\tt RNASSD} software~\cite{AnFeHu2004}. 
Here, we supplement this study by comparing the global sampling approach implemented by {\tt CFGRNAD} with {\tt NUPACK}~\cite{Zadeh:2011uq}. 

\begin{figure}[t]
  {\centering
  \includegraphics[width=.9\linewidth]{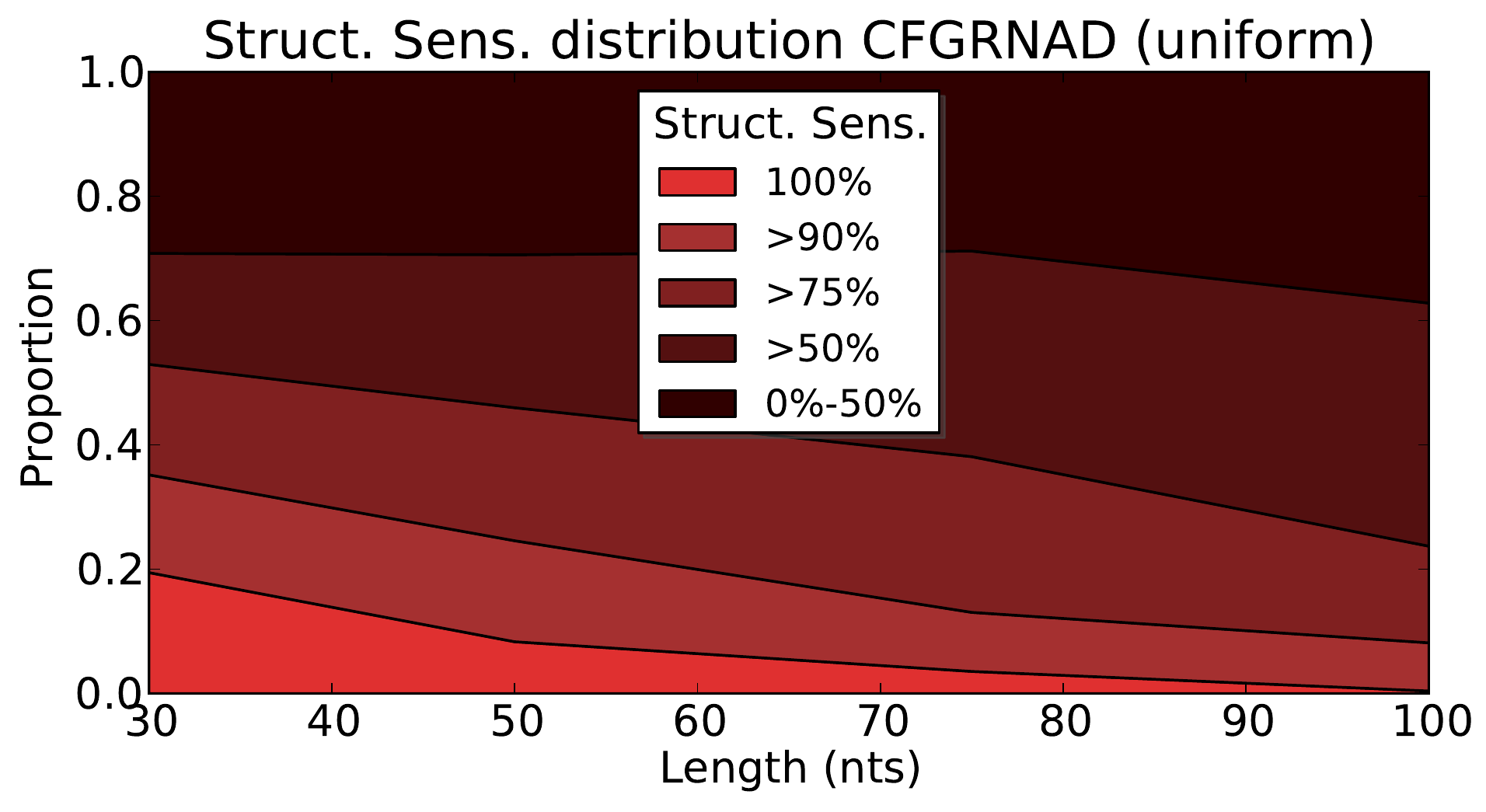}
  \includegraphics[width=.9\linewidth]{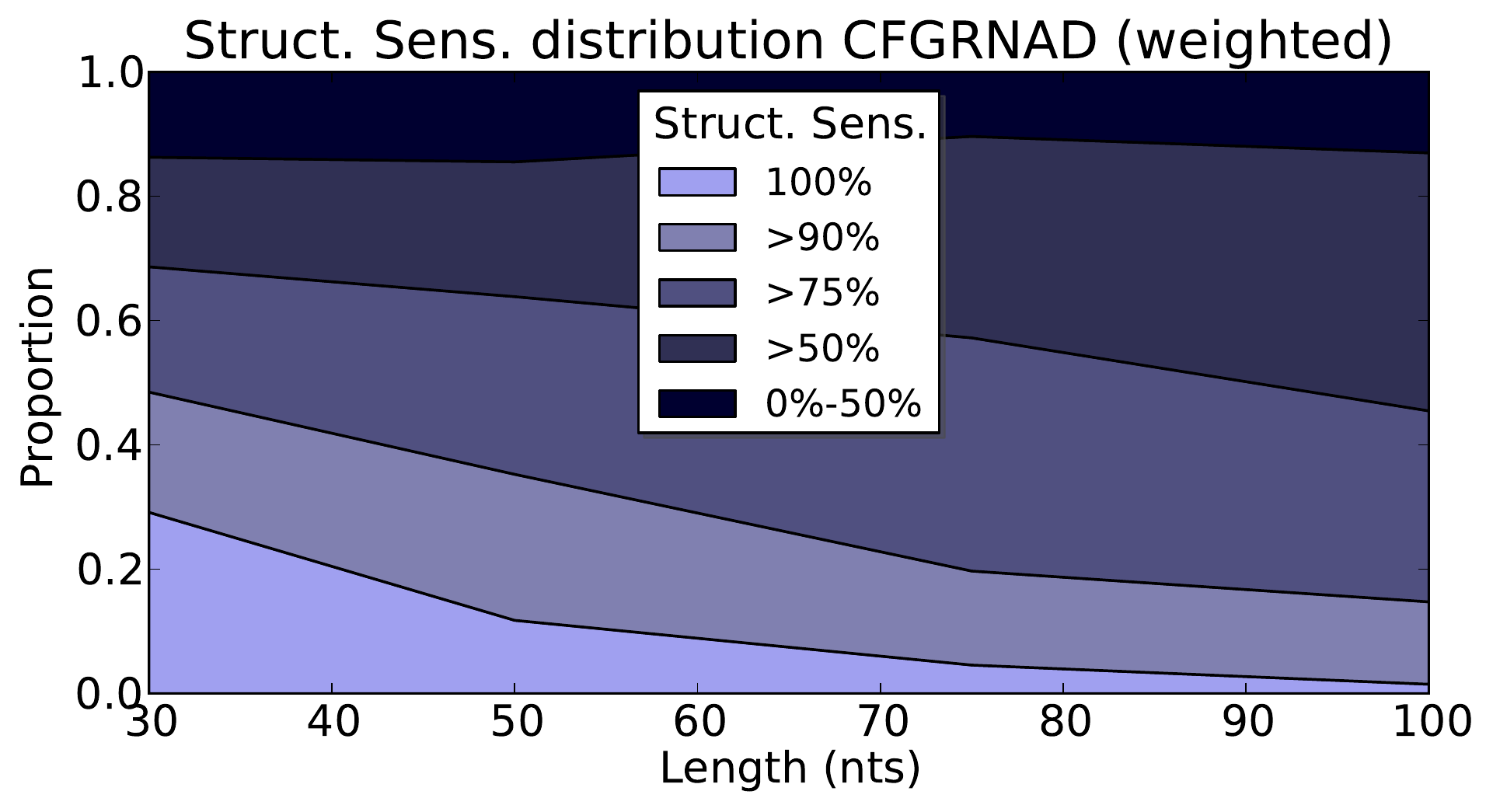}
  \includegraphics[width=.9\linewidth]{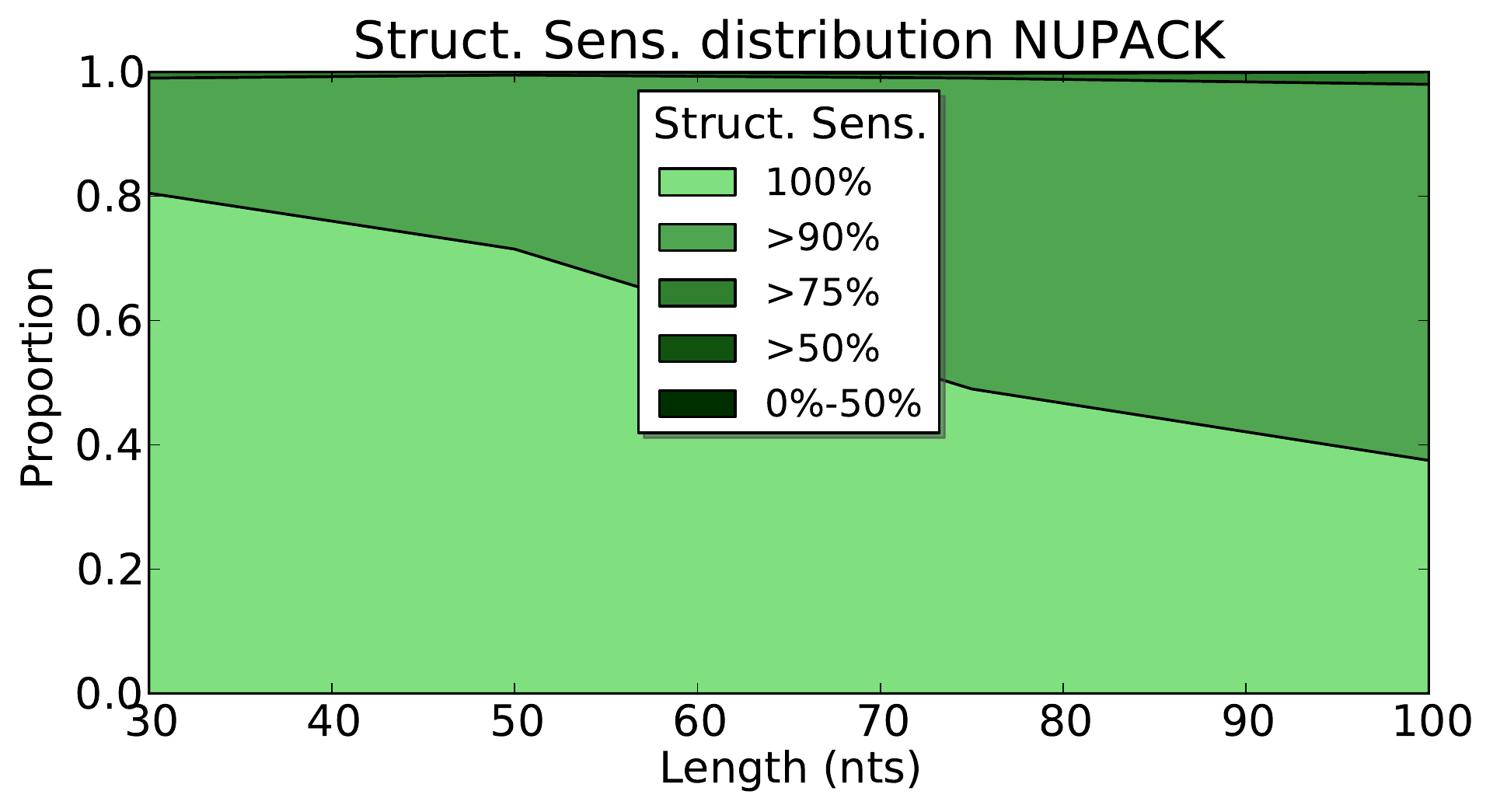}\\}

  \caption{Structural sensitivity distribution for sequences designed using  {\tt CFGRNAD} 
(uniform distribution, top), {\tt CFGRNAD} 
(weighted distribution, middle) and {\tt NUPACK} (bottom).\label{fig:distrib}}
\end{figure}
Our benchmark consists in sets of 20 random target secondary structures for each length in $[30,50,75,100]$, generated as described in Reinharz \emph{et al}~\cite{Reinharz2013}. For each target structure, we designed a set of 20 design candidates using {\tt NUPACK} (random initial seed + Turner 1999 energy model). 
For difficult designs, {\tt NUPACK} may produce sequences whose MFE-predicted structure does not exactly match the target structure, yet share many base-pairs with it. In order to assess the quality of the designed sequence, we define the structural sensitivity of a sequence $w$ with respect to a target $ \Target$ as:
\begin{equation} {\sf Sen}(w,\Target) = \frac{\left|\{(i,j)\;|\;(i,j)\in \Target \text{ and }(i,j)\in \Target_w\}\right|}{|w|} \end{equation}
where $\Target_w$ is the MFE structure for $w$, predicted using {\tt RNAfold} (Turner 1999 energy model).
Then we ran our software {\tt CFGRNAD} for a total runtime which is equal to that of {\tt NUPACK}. We  include in both measured runtimes the computation of structural sensitivities. This property seems essential for a fair comparison, as {\tt CFGRNAD} produces much more candidate sequences than its competitor. Furthermore, we emphasize that the energy model (Turner 1999) used to score a design is the same as the one used by {\tt NUPACK} in its optimization, but differs from that of {\tt CFGRNAD} (modified Turner 2004).

An analysis of the distribution of structural sensitivities, shown in Figure~\ref{fig:distrib}, confirms the positive impact of weights on the quality of {\tt CFGRNAD} designs. The relative proportions of successful designs remains highly favorable to {\tt NUPACK}. For the latter, the average probabilities of complete success ranged from 80\% down to 40\% for sequence lengths between 30 and 100 nts, and 97\% of refolded designs shared at least 90\% of base-pairs with their target. By contrast, {\tt CFGRNAD} only achieved 30\%-2\% for complete success and 50-18\% for >90\% sensitivity. 

\begin{figure}[t]
  {\centering
  \includegraphics[width=.98\linewidth]{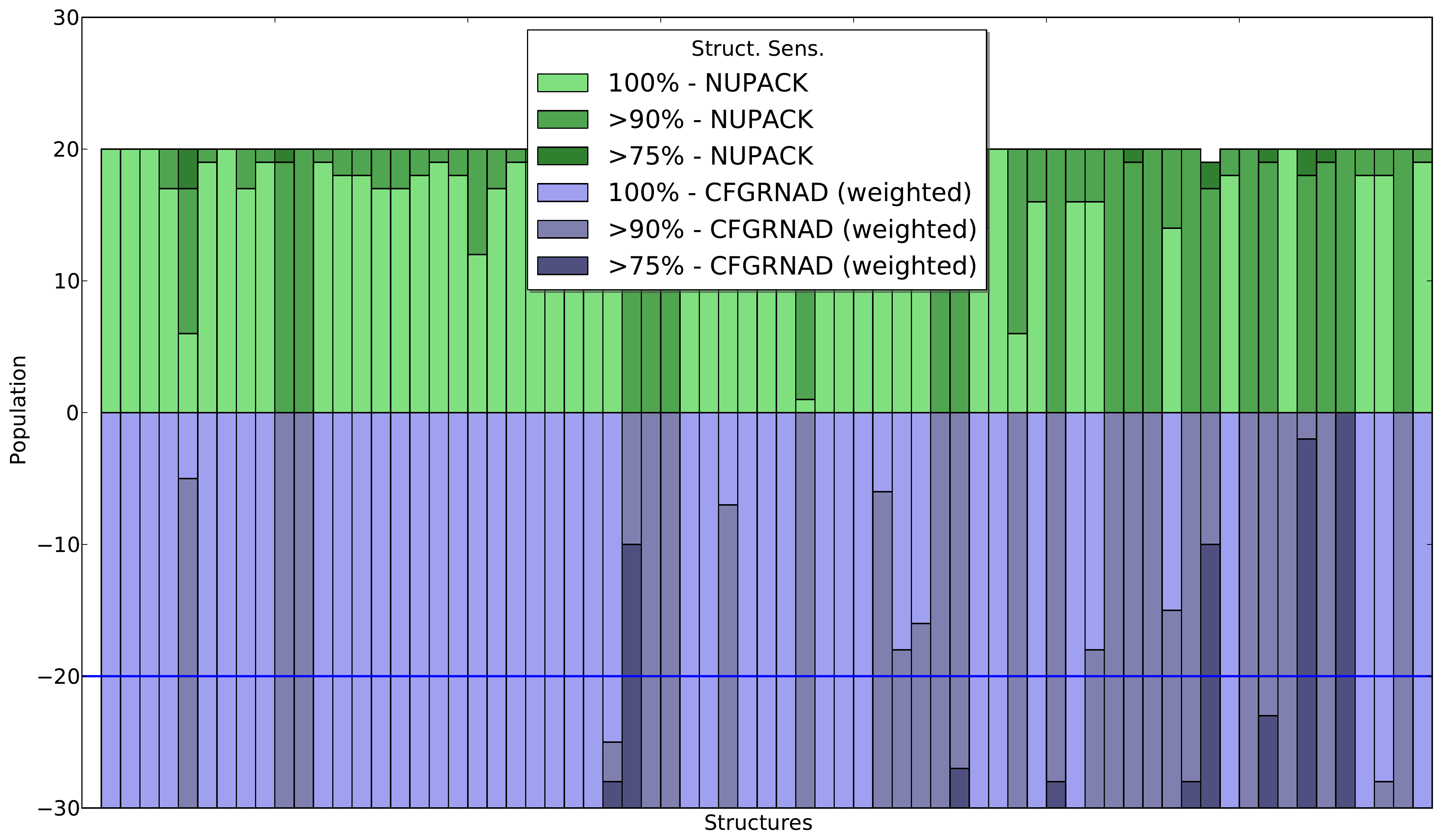}\\}

  \caption{Raw population and structural sensitivity of candidate sequences generated by {\tt NUPACK} (Green, upward) and {\tt CFGRNAD} (Blue, downward) for target structures of increasing lengths, clustered according to their structural sensitivity. \label{fig:population}}
\end{figure}

However, the linear time complexity of {\tt CFGRNAD} allows for the generation of larger sets of candidates, even when the cubic-time  {\tt RNAfold} is included in the postprocessing. 
For instance, using {\tt NUPACK} to generate 20 candidates for a structure of 100 nts may require as much as 6 800 seconds on a standard laptop, allowing for the generation and evaluation of 197 000 {\tt CFGRNAD} candidates. As shown in Figure~\ref{fig:population}, this larger number of candidates produced by {\tt CFGRNAD} in a given time apparently compensates its lower probability of success and, with very few notable exceptions, the raw numbers of perfect, good and acceptable designs produced by  {\tt CFGRNAD} seems to exceed that of {\tt NUPACK}. One should however remain cautious before extrapolating this claim for larger RNAs, since the probability of producing perfect designs through random generation is expected to decrease exponentially with the length. 

\section{Conclusion}
\label{conclusion}

In this work, we have proposed a novel approach based on language-theoretical tools to perform a rational design of RNA under a biologically-motivated set of constraints.
Namely, we showed that large set of diverse constraints could be addressed within a unifying framework, leading to an algorithm whose time and space requirements scale linearly with the
length of designed secondary structures. Furthermore, such a framework paves the road for novel extensions, and our program is currently the only available tool to perform RNA design 
with forced/forbidden motifs. Such a feature has proven its utility in the design of sequences that have been tested \emph{in vitro}. 

A first extension of this work may assess the impact of, possibly stringent, constraints on the probability of actually folding into the MFE. 
In the absence of such constraints, it was shown that a substantial proportion of sequences designed using a weighted random generation scheme (also denoted as \emph{global sampling strategy})
actually folds back \emph{in silico} into the target structure~\cite{Levin:2012kx,Reinharz2013}. The success rate is sufficiently high to make the approach competitive against the alternatives based on various heuristics or exponential-time exact resolution, especially when additional design goals are captured~\cite{Reinharz2013}.
However, certain constraints may possibly have a drastic, non-trivial, effect on the success rate, for instance by preventing motifs favored by stable folds from forming (e.g. forbidding $\Gb\Cb,\Cb\Gb,\Cb\Cb$  motifs would forbid any occurrence of a $\frac{\Gb\Cb}{\Gb\Cb}$ stacking pairs). Such an effect might also be interpreted in the light of evolution, as the low designability (i.e. evolutionary accessibility) of experimental secondary structures may indicate a strong selective pressure, and suggest an essential position in the RNA regulatory network. 

Another illustration of the flexibility of our approach is the inclusion of structural (possibly non-canonical) motifs, which could be expressed as multi-strand (or gapped) motifs, contextualized by the target secondary structure. These may be prescribed either at a specific location in the sequence, or anywhere, depending on the intended application. The design of sequences of variable lengths may also constitute a natural extension, at the cost of more demanding algorithms. Future developments will aim at capturing more sophisticated free-energy models, such as the full Turner model, or crossing interactions (pseudoknots), relying on other language-theoretical formalisms such as multiple grammars.

\section*{Acknowledgment}

The authors would like to thank James Regis (LIX, Polytechnique)
for technical help with the webserver infrastructure.

\bibliographystyle{splncs03}
\bibliography{design}

\end{document}